\documentclass[11pt,a4paper]{article}

\usepackage{amsmath}
\usepackage{amssymb}
\usepackage{graphicx,subfigure}
\author{Atsushi Iwasaki\thanks{Fukuoka Institute of Technology\ \  e-mail: a-iwasaki@fit.ac.jp}}
\title{Analysis of NIST SP800-22 focusing on\\ randomness of  each sequence}
\begin{document}

\maketitle
\abstract{
NIST SP800-22 is a randomness test set applied for a set of sequences.
Although SP800-22 widely used, a rational criterion throughout all test items has not been shown.
The main reason is that the dependency of test items has not been perfectly clear.
In this paper, a certain scalar is computed for each sequence throughout all test items and make the histogram of the scalar.
By comparing the histogram and the theoretical distribution under some assumptions, the dependency is 
visually shown.
In addition, an algorithmic method to derive ``minimum set'' using the histogram is proposed.
}

\section{Introduction}
Random number sequence is used in many field and, in particular, it is indispensable for cryptography.
In cryptography, high randomness is demanded.
Thus, random number sequence and its generator used in such field are evaluated from many view points.
Randomness test is one of such view points.
It is a hypothesis test applied for sequence and the null hypothesis is that given sequence is ideally random. 
It is applicable for any sequence irrespective of its generator.
It is, therefore, very useful for the evaluation although it gives only numerical result.

Countless randomness tests can be considered, but it is impossible to perform all tests.
Then, only selected tests are performed.
NIST SP800-22 \cite{NIST} is one of such test sets.
The resent version of SP800-22 consists of 188 test items of 15 kinds and  is widely used.

Notwithstanding, SP800-22, even the current version, has some problems.
One of the problems is that a rational criterion throughout all test items has not been shown.
The results of different test items are not mutually independent and the fact is verified by some researches such as Ref. \cite{Lihua}.
The fact makes it difficult to create a rational criterion. 
Indeed, there are some researches such as Ref. \cite{Okutomi} which proposed a criterion, but they assumed that test items are mutually independent and so the criterion is not rational.

In order to solve the problem, study on the dependency is important.
SP800-22 is applied for a set of sequences and p-value is computed for each combination of a sequence and a test item.  
If the sequence is ideally random, the p-value approximately follows the uniform distribution on the interval [0,1].
Almost prior numerical researches focus on correlation between two test items based on p-values or numbers of test items which sequences passed, but such approach has not achieved sufficient knowledge which can make creating a rational criterion easy. 
Then, in this paper, we propose another new approach.

This paper is constructed as follows:
In Section 2, an approach is introduced.
In Section 3, a simple example of the approach is shown.
In Section 4, an algorithmic method to derive ``minimum set'' using the example is proposed.
Finally, we conclude this paper.


\section{Proposed approach}

NIST SP800-22 is applied for a set of sequences.
As the null hypothesis, it is assumed that the sequences are ideally random and mutually independent.
Assume that a set is consists of $m$ sequences and the length of each sequence is $n$-bit.
Each test included in SP800-22 except Random Excursion Test and Random Excursion Variant Test is applied for all the $m$ sequences. 
Random Excursion Test and Random Excursion Variant Test are applied only for sequences which satisfy some conditions.
Since the specification of the two kind of tests is not good for the approach proposed in this paper, we exclude the two kinds of test, 26 items, and consider only 13 kinds, 162 items.
The same exclusion was also done in Ref. \cite{Okutomi}.

Assume that $p_{i,j}$ means the p-value which test item $i$ computes for sequence $X_j$ ($i=1,2,\cdots,162,\ j=1,2,\cdots,m$).
The approach proposed in this paper is as follows:
\begin{enumerate}
\item For each sequence $X_j$, compute a certain scalar value as
\[q_j=q(p_{1,j},p_{2,j},\cdots,p_{162,j}).\]
\item Make the histogram of $q_1,q_2,\cdots,q_m$ and compare it with theoretical distribution computed under the assumption that all test items are mutually independent.
\end{enumerate}
This approach has the following advantages:
\begin{itemize}
\item If a simple function is used as $q$, the theoretical distribution with the assumption of independency can be easily derived.
\item Since only one sequence $X_j$ is needed for computing $q_j$ and $X_1,X_2,\cdots,X_m$ are mutually independent by the null hypothesis, $q_1,q_2,\cdots,q_m$ can be regarded as mutually independent.
\end{itemize}
Then, we can easily observe the dependency of the test items as the deviation between the histogram of  $q_1,q_2,\cdots,q_m$ and the theoretical distribution with the assumption of independency.


\section{Example of the proposed approach}
As an example, we define $q$ as
\[q_j:=\frac{p_{1,j},p_{2,j},\cdots,p_{162,j}}{162},\]
and make some experiments.
By the central limit theorem, $q_j$ can be approximately regarded as a random variable following the normal distribution whose average and standard deviation are $\frac{1}{2}$ and $\frac{1}{\sqrt{162\times12}}$ respectively, if the null hypothesis and the assumption that the test items are mutually independent are true.

\begin{figure}
    \centering\subfigure[$n=10^4$, $m=10^6$.]{\includegraphics[scale=0.25]{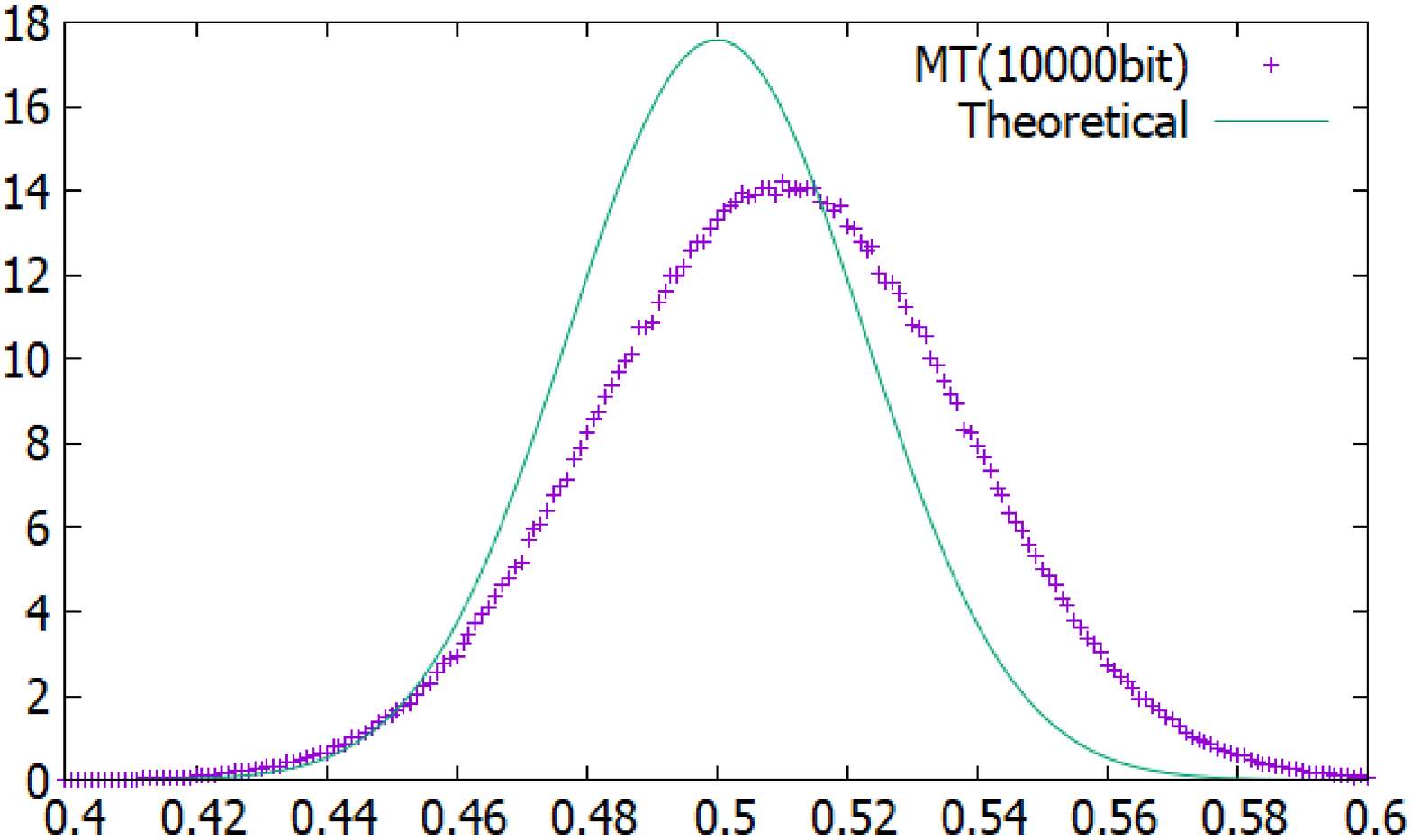}}
    \centering\subfigure[$n=10^5$, $m=10^6$.]{\includegraphics[scale=0.25]{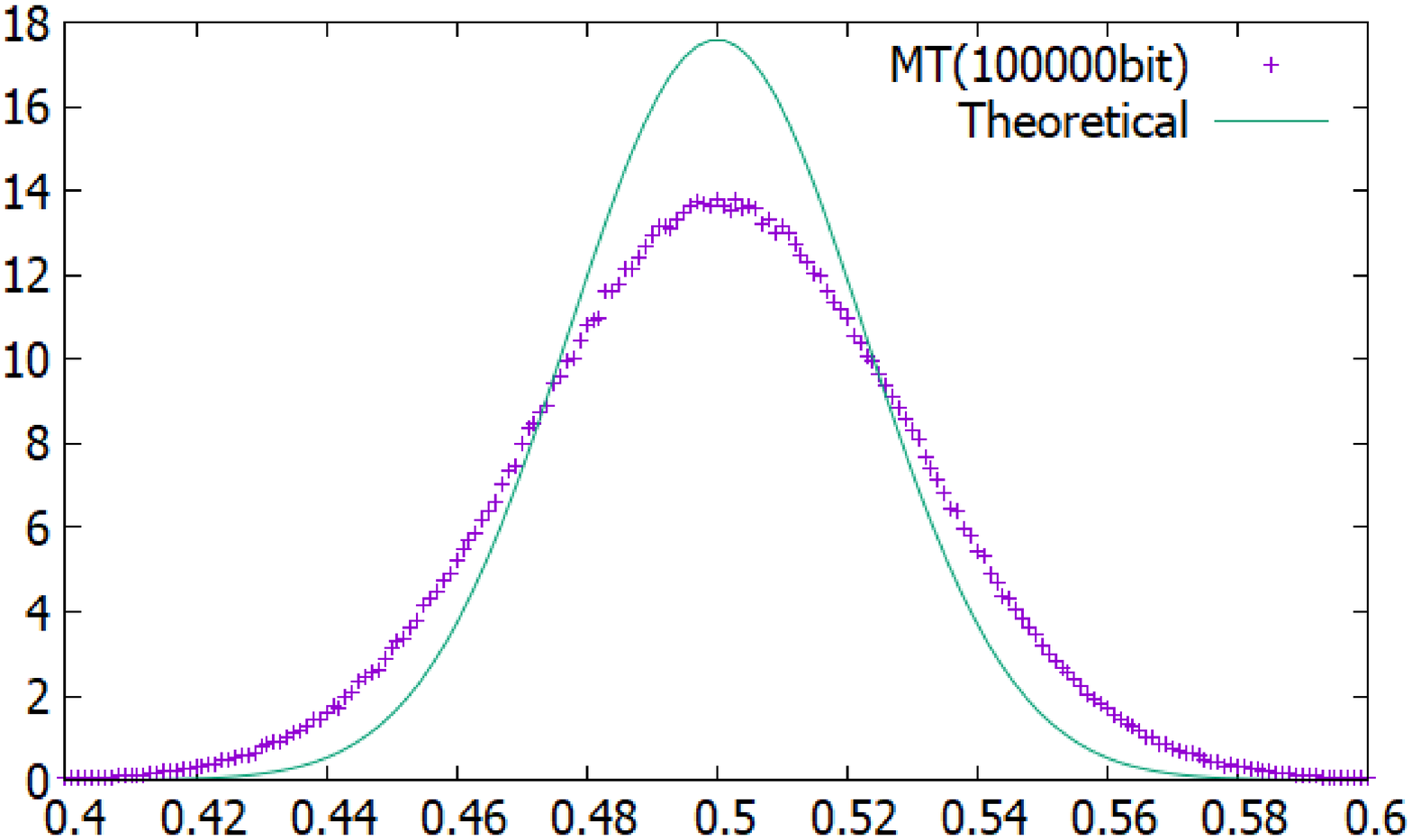}}
    \centering\subfigure[$n=10^6$, $m=10^6$.]{\includegraphics[scale=0.25]{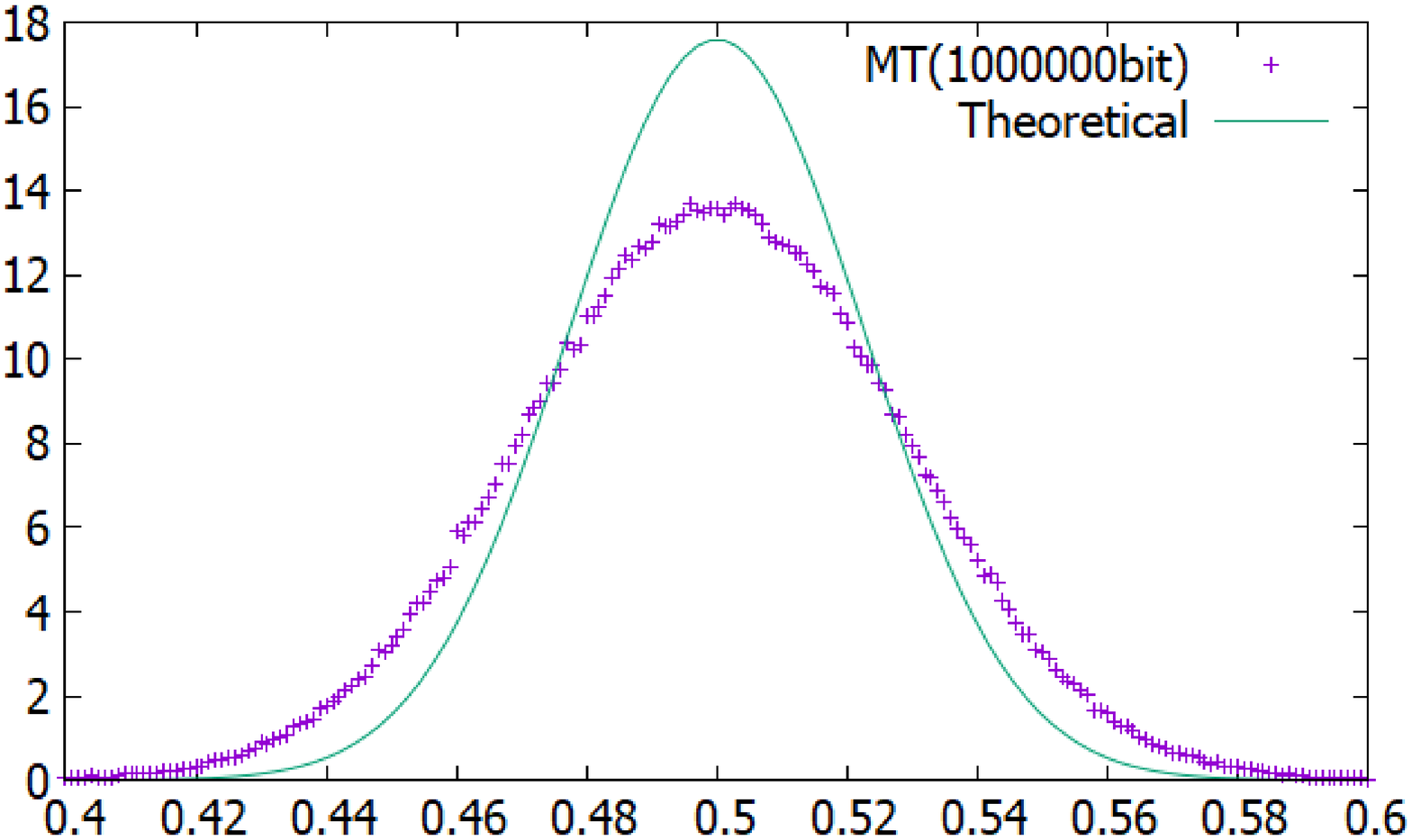}}
    \centering\subfigure[$n=10^7$, $m=10^5$.]{\includegraphics[scale=0.25]{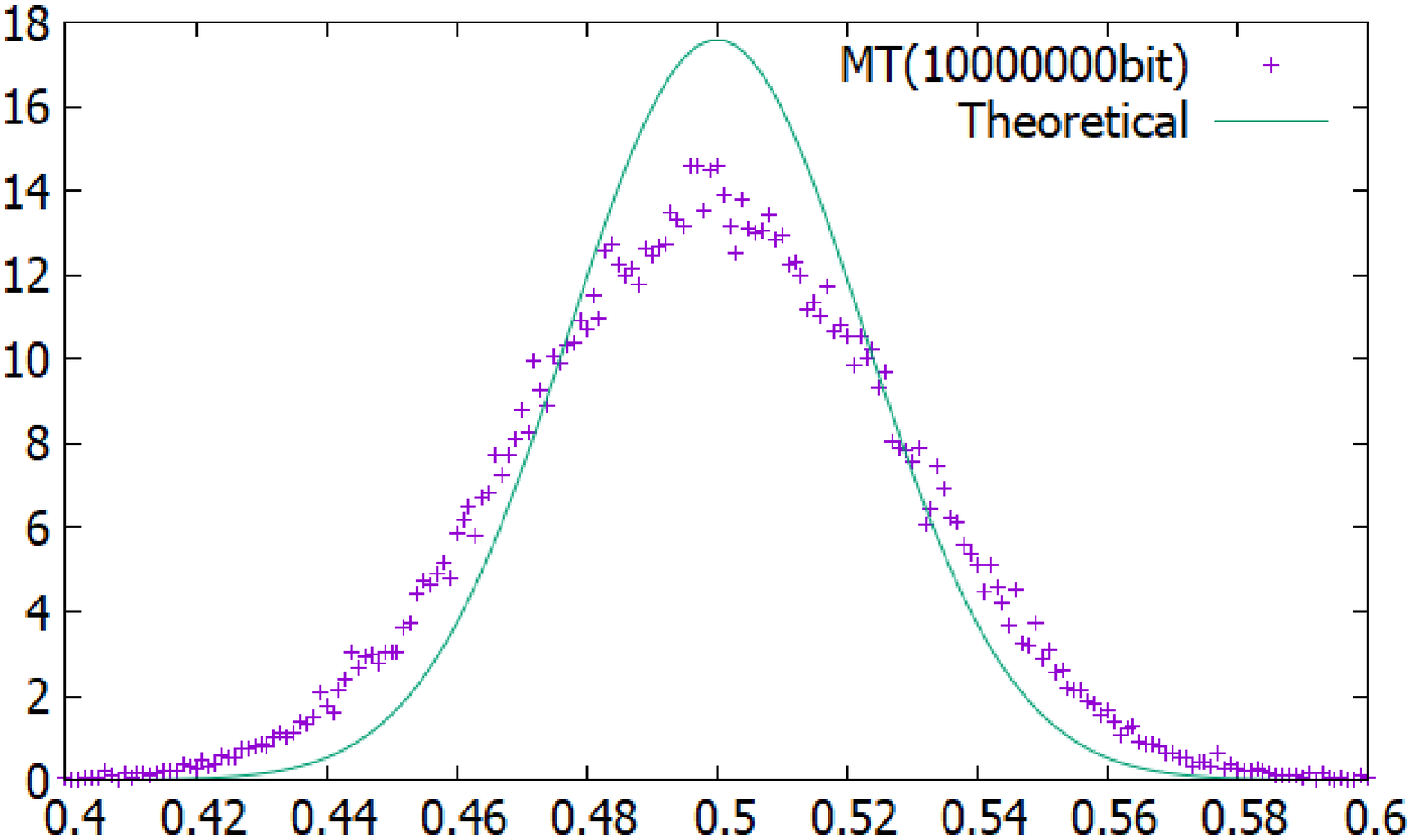}}
\caption{Comparing the histogram of $q_1,q_2,\cdots,q_m$ made from sequences generated by Mersenne twister with the theoretical distribution.}
\label{MT}
\end{figure}

\begin{figure}
    \centering\subfigure[$n=10^4$, $m=10^6$.]{\includegraphics[scale=0.25]{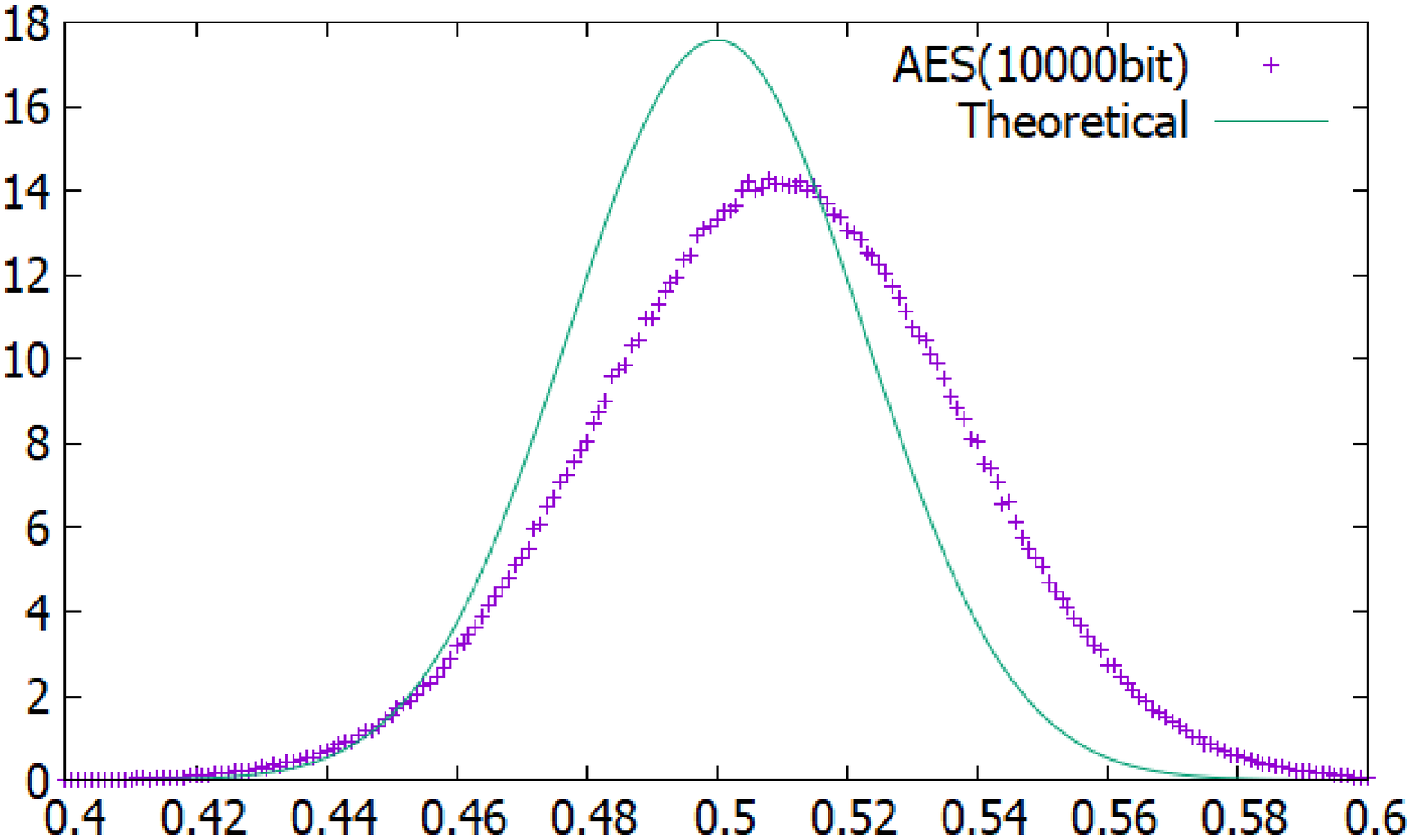}}
    \centering\subfigure[$n=10^5$, $m=10^6$.]{\includegraphics[scale=0.25]{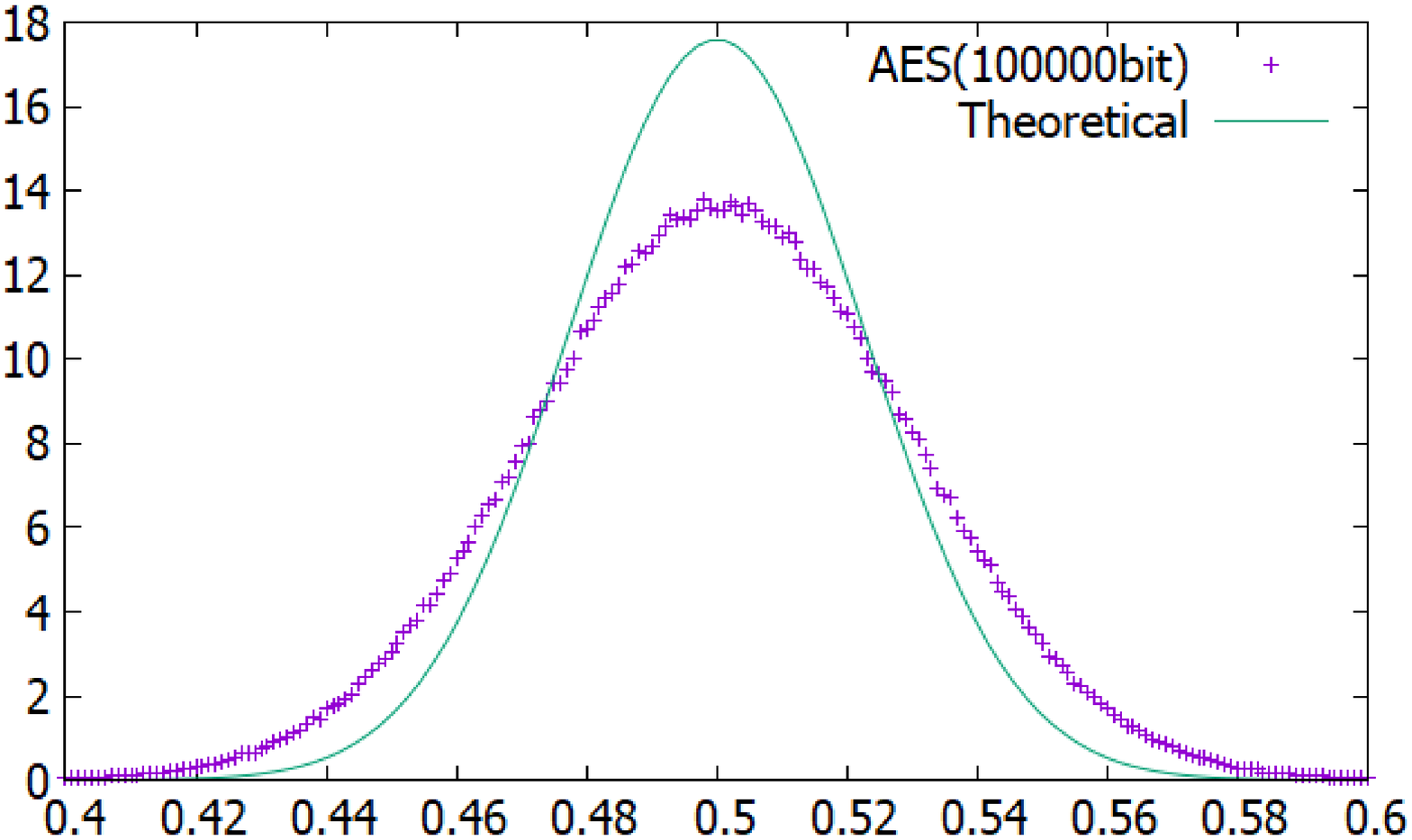}}
    \centering\subfigure[$n=10^6$, $m=10^6$.]{\includegraphics[scale=0.25]{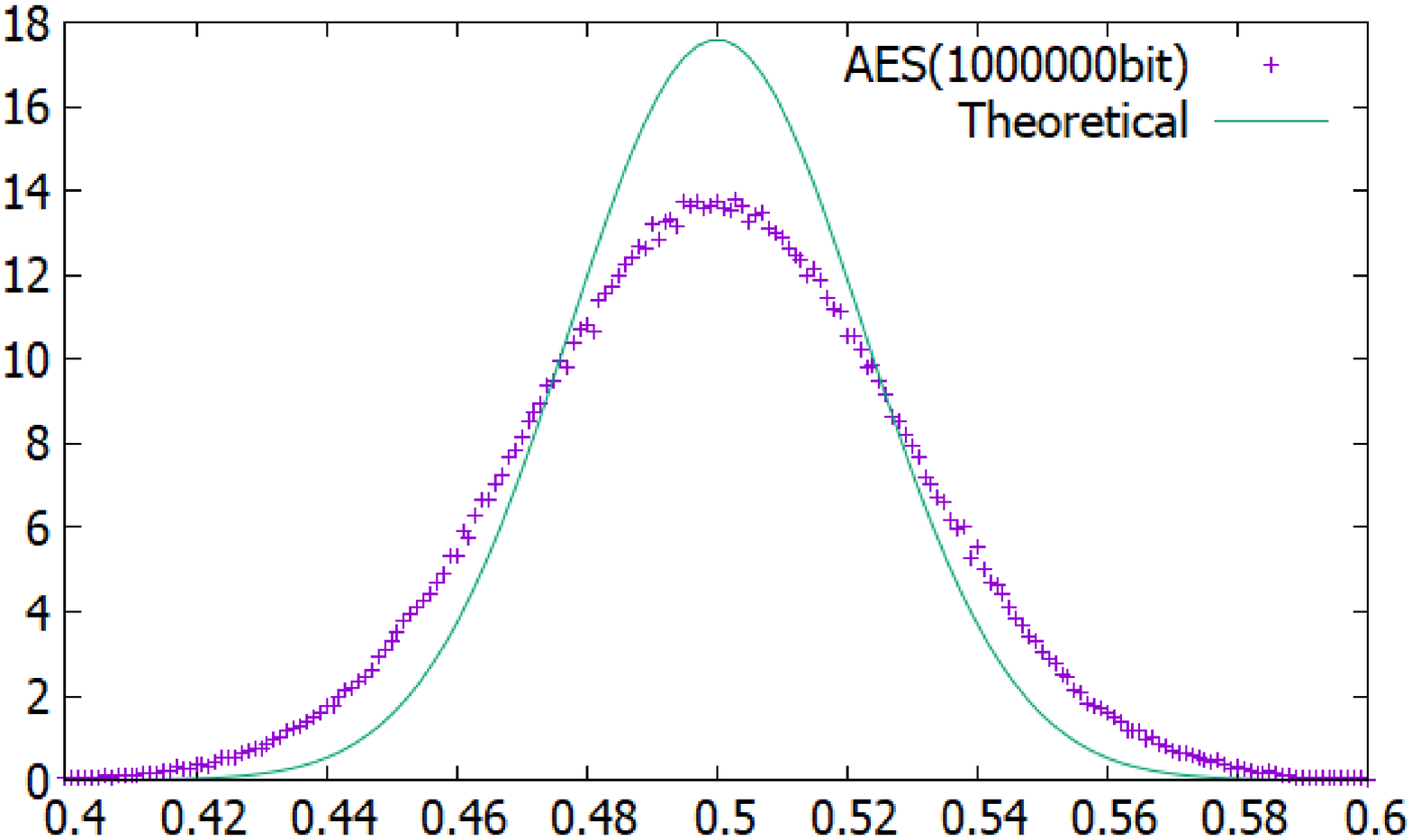}}
    \centering\subfigure[$n=10^7$, $m=10^5$.]{\includegraphics[scale=0.25]{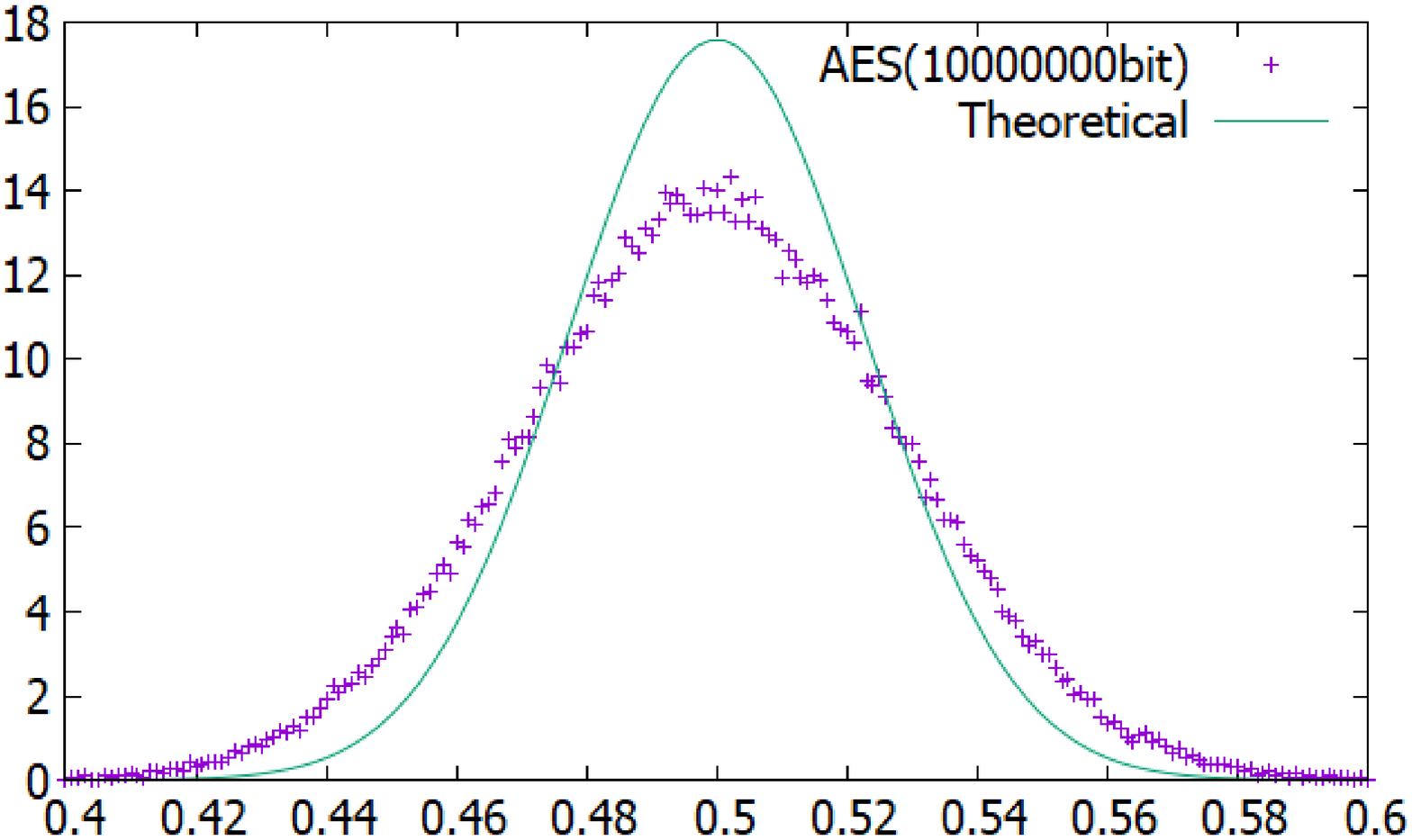}}
\caption{Comparing the histogram of $q_1,q_2,\cdots,q_m$ made from sequences generated by AES with the theoretical distribution.}
\label{AES}
\end{figure}

Fig. \ref{MT} and \ref{AES} show comparison between the normal distribution and the histogram of $q_1,q_2,\cdots,q_m$ made from sequences generated by Mersenne twister \cite{MT} and AES \cite{AES}, respectively.
The histograms are wider than the normal distribution.
 These results suggest that there is a strong positive correlation among the test items.

\begin{figure}
    \centering\subfigure[$n=10^4$, $m=10^6$.]{\includegraphics[scale=0.25]{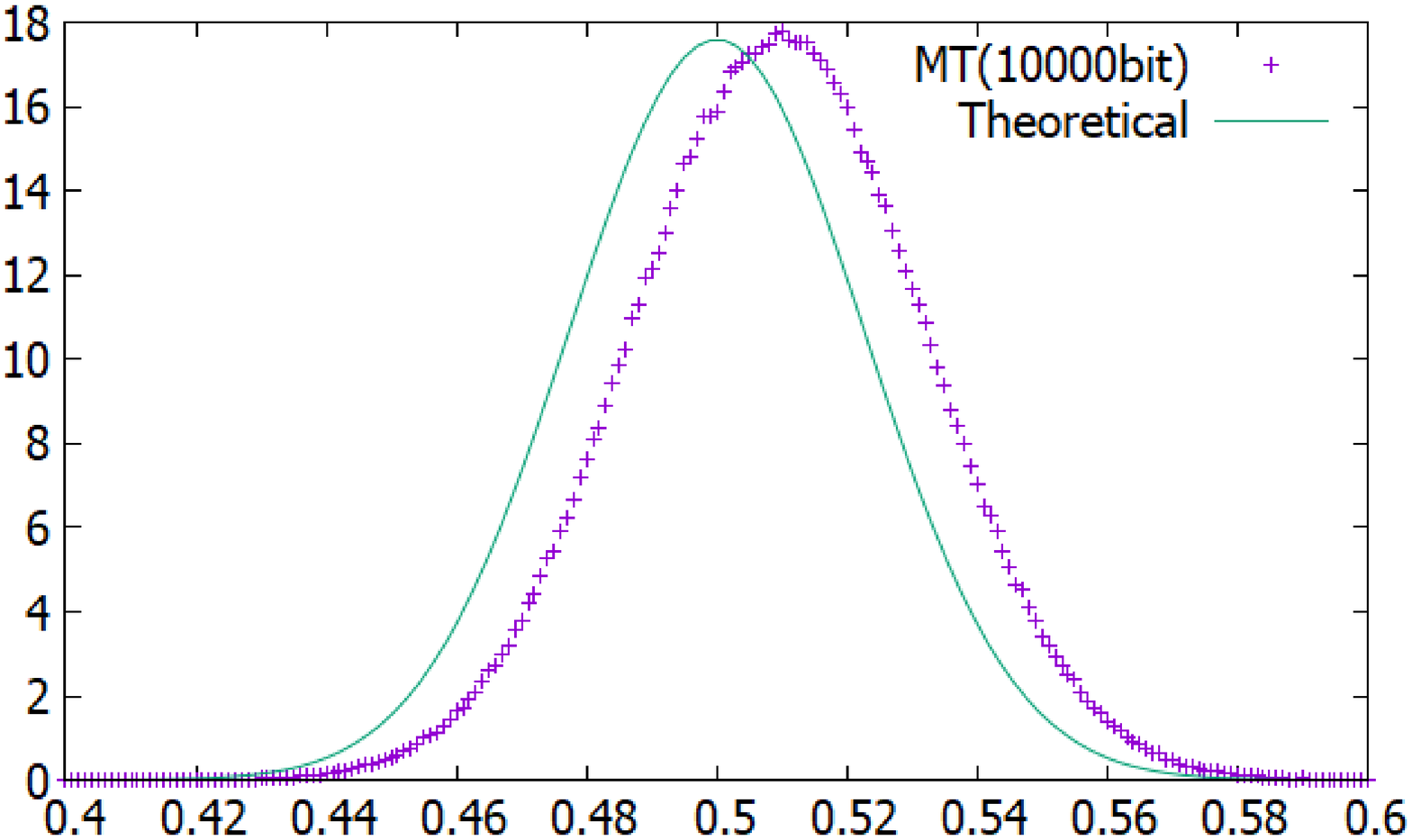}}
    \centering\subfigure[$n=10^5$, $m=10^6$.]{\includegraphics[scale=0.25]{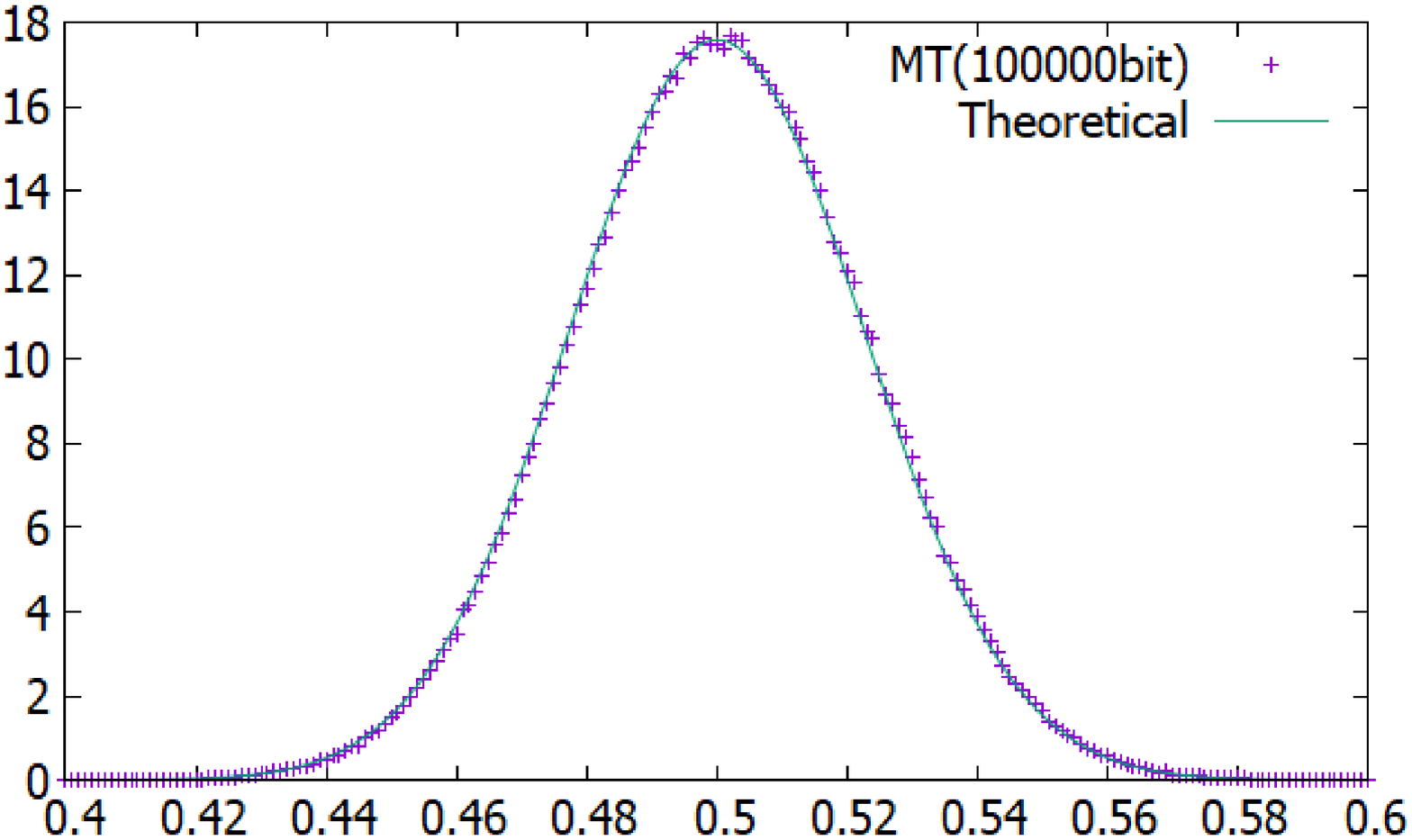}}
    \centering\subfigure[$n=10^6$, $m=10^6$.]{\includegraphics[scale=0.25]{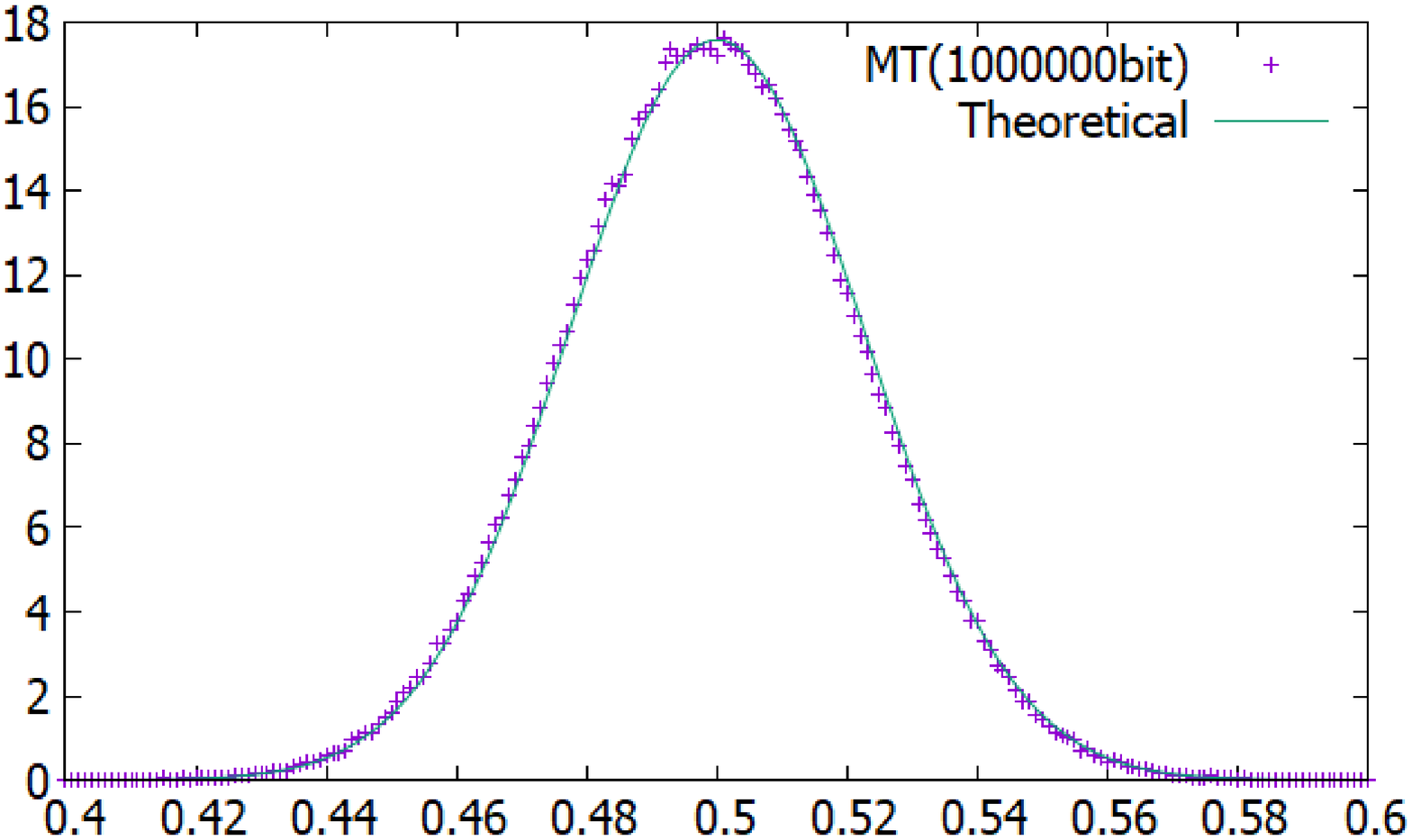}}
    \centering\subfigure[$n=10^7$, $m=10^5$.]{\includegraphics[scale=0.25]{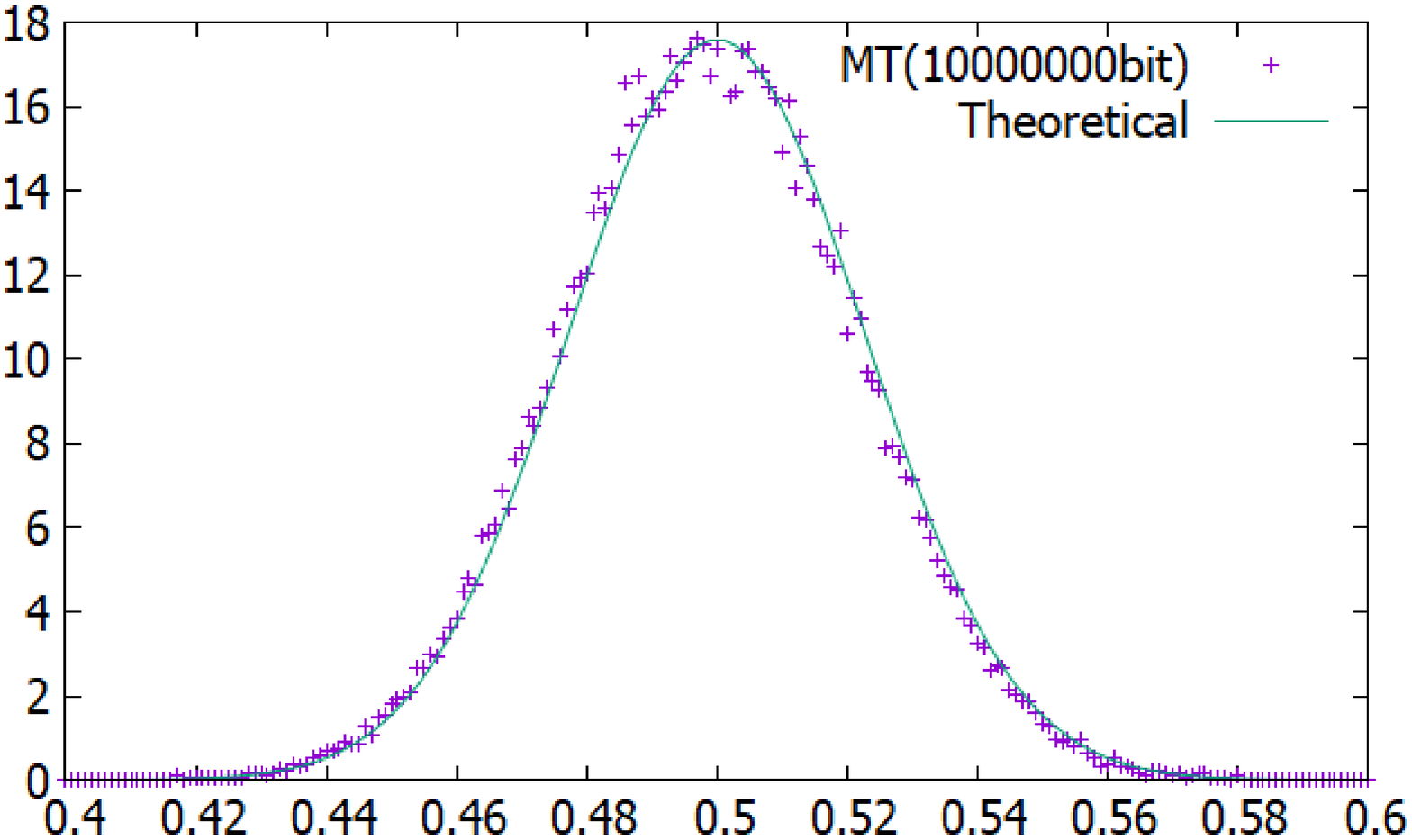}}
\caption{Comparing the histogram of $\tilde{q}_1,\tilde{q}_2,\cdots,\tilde{q}_m$ made from sequences generated by Mersenne twister with the theoretical distribution.}
\label{MT_scramble}
\end{figure}

\begin{figure}
    \centering\subfigure[$n=10^4$, $m=10^6$.]{\includegraphics[scale=0.25]{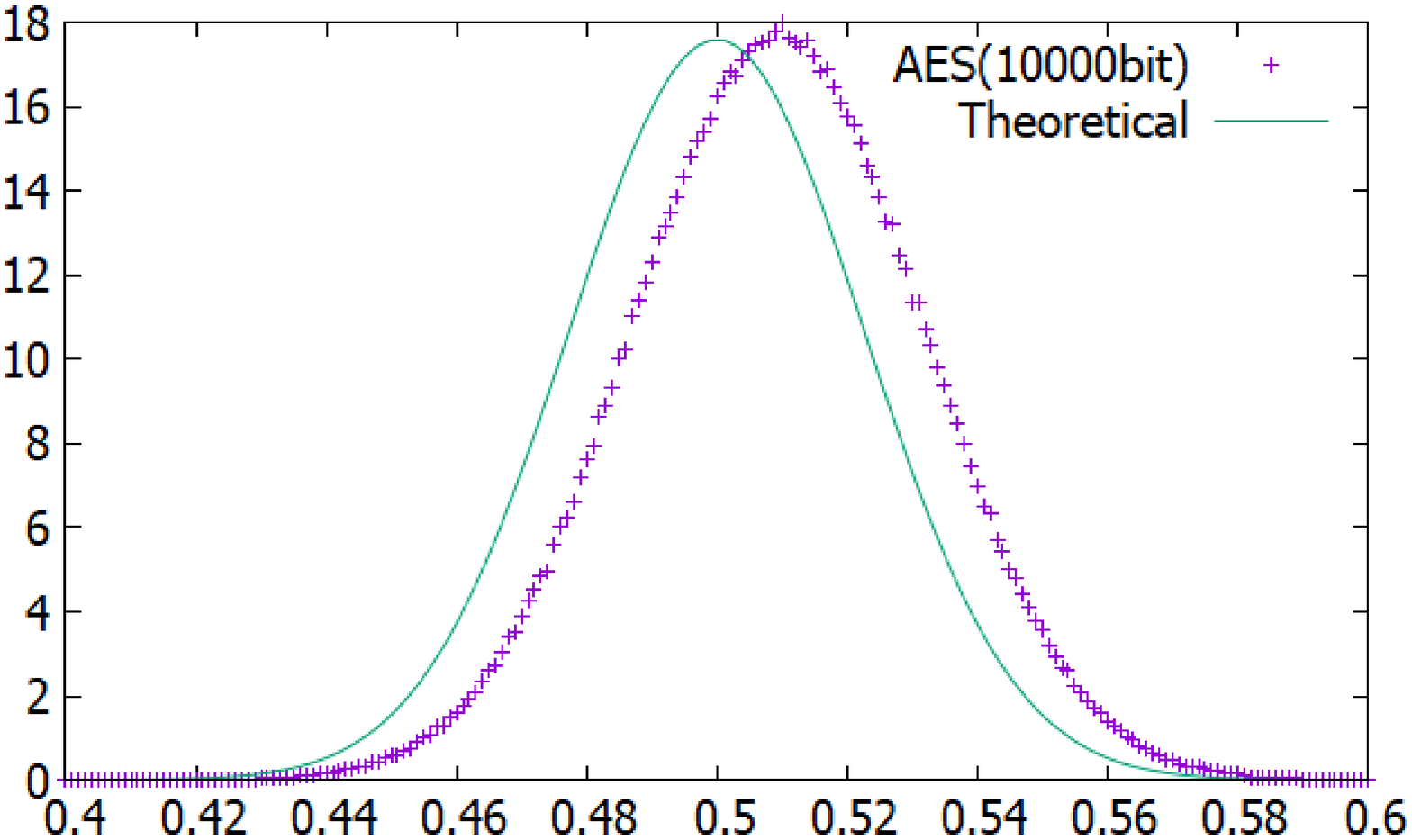}}
    \centering\subfigure[$n=10^5$, $m=10^6$.]{\includegraphics[scale=0.25]{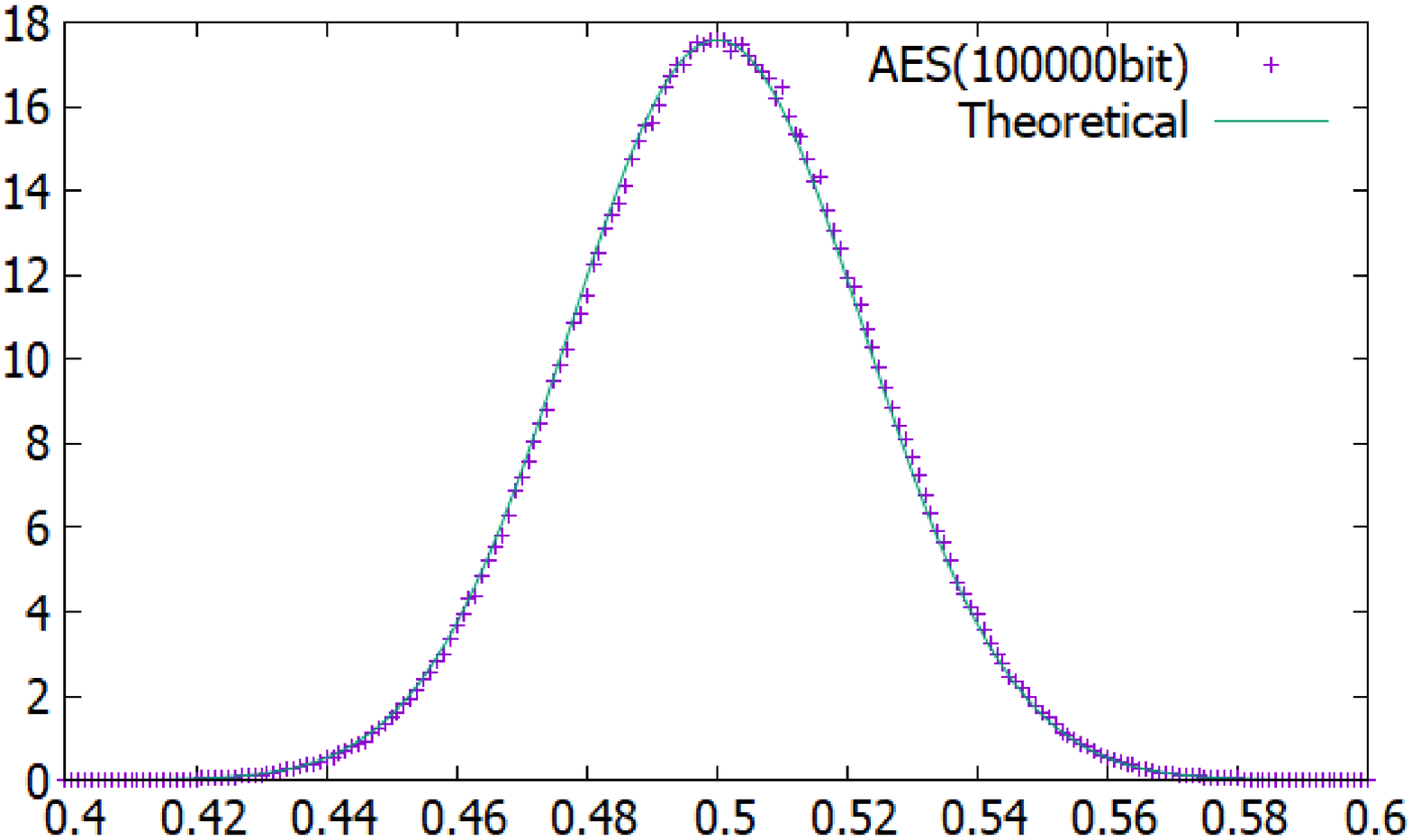}}
    \centering\subfigure[$n=10^6$, $m=10^6$.]{\includegraphics[scale=0.25]{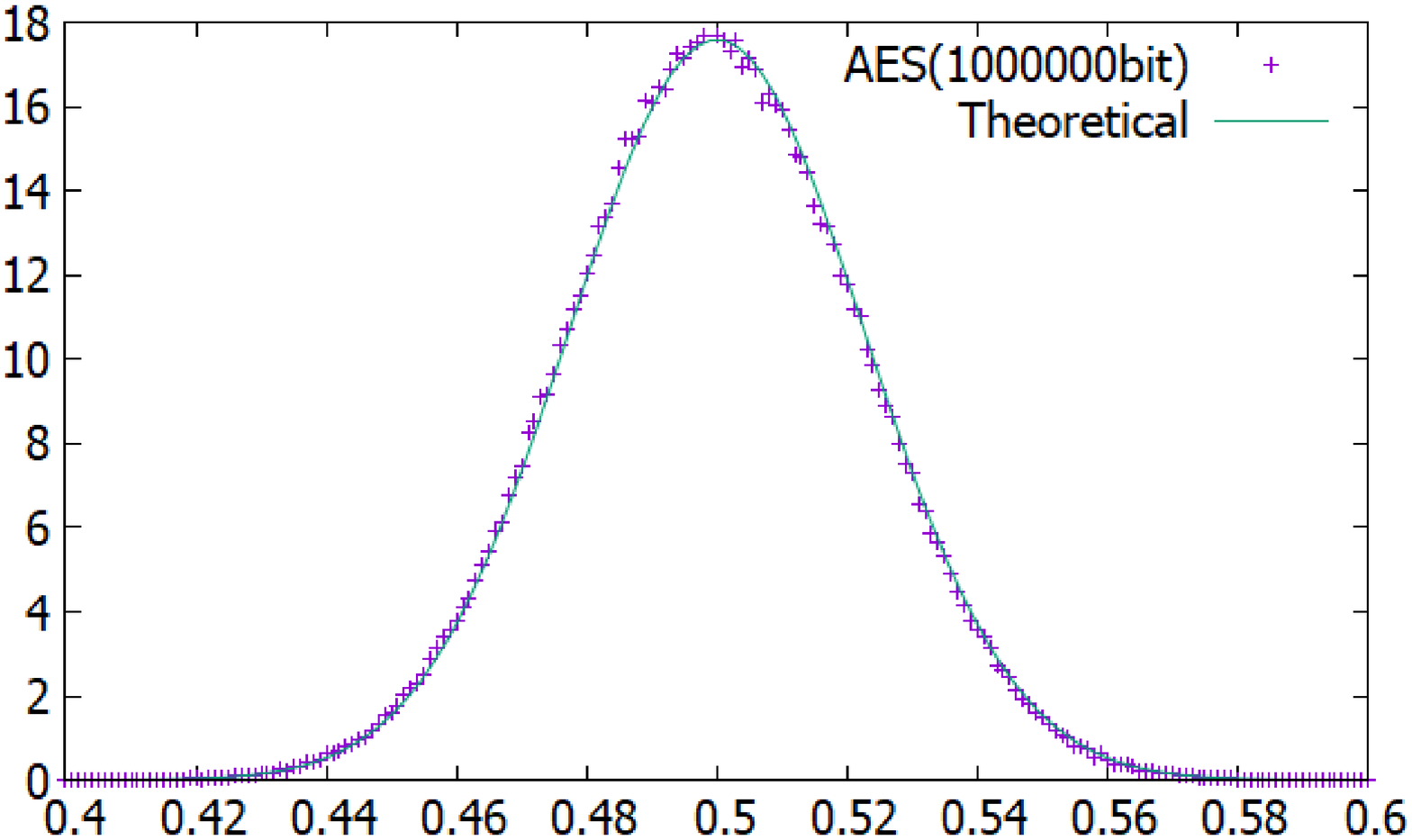}}
    \centering\subfigure[$n=10^7$, $m=10^5$.]{\includegraphics[scale=0.25]{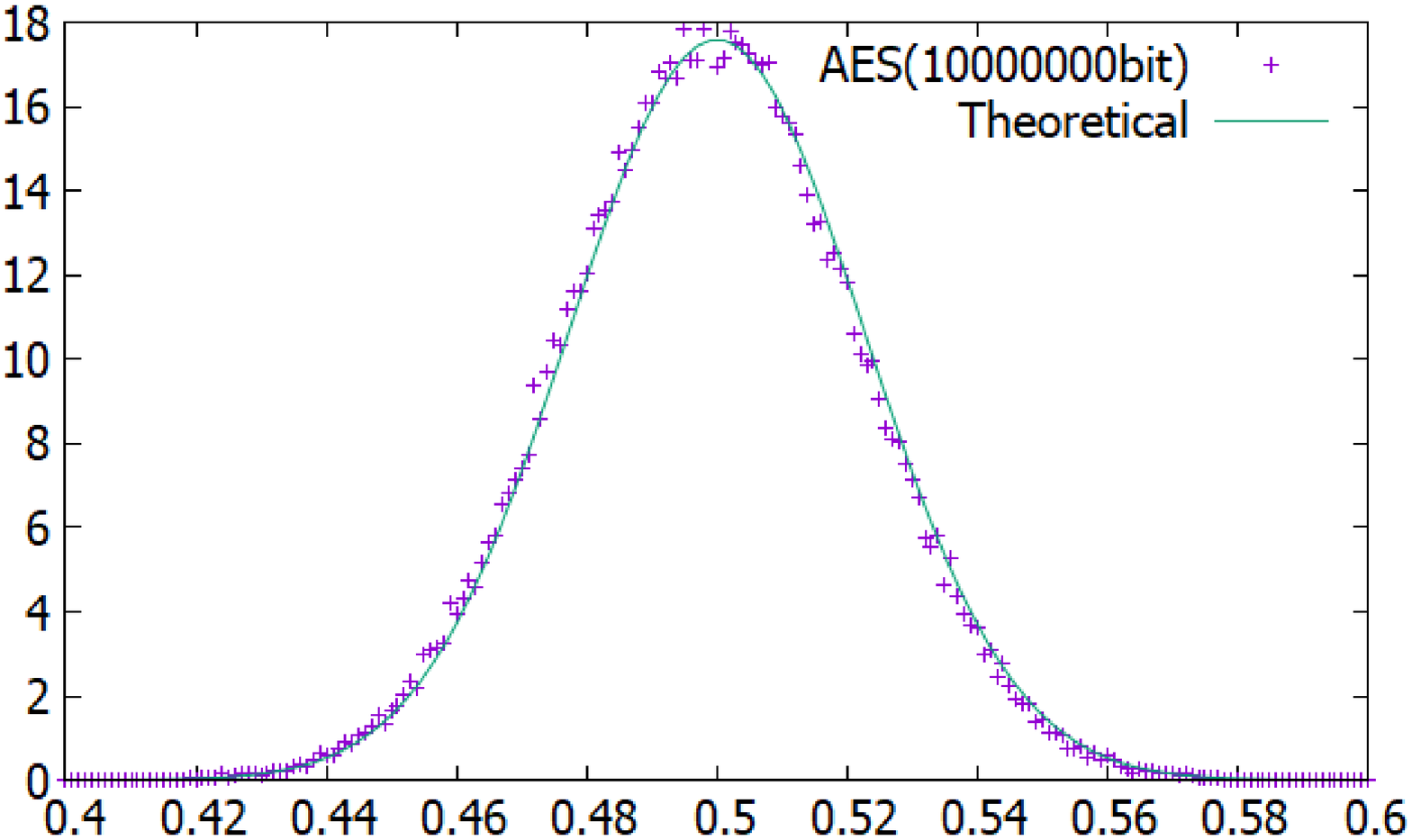}}
\caption{Comparing the histogram of $\tilde{q}_1,\tilde{q}_2,\cdots,\tilde{q}_m$ made from sequences generated by AES with the theoretical distribution.}
\label{AES_scramble}
\end{figure}

In order to verify the suggestion, we make another experiment.
Exactly speaking, p-values of each test included in SP800-22 do not follow  the uniform distribution on [0,1].
One of the reasons is that the sequence length is finite in real situation although the distribution of p-values is expected to approach the uniform distribution as $n\to\infty$.
In addition, it is reported that design of some tests include some mistakes. (See Ref. \cite{Pareschi}. as an example.)
If the fact is the reason of the deviations shown in Fig. \ref{MT} and \ref{AES}, the histogram of $\tilde{q}_1,\tilde{q}_2,\cdots,\tilde{q}_m$ also deviates from the normal distribution, where $\tilde{q}_j$ is defined as
\[\tilde{q}_j:=\frac{1}{162}\sum_{i=1}^{162}p_{i,(j+i\mod m)}.\]
Fig. \ref{MT_scramble} and \ref{AES_scramble} shows the results of Mersenne twister and AES, respectively. 
The widths of the histograms of  $\tilde{q}_1,\tilde{q}_2,\cdots,\tilde{q}_m$ is near the normal distribution.

By these results, we can conclude that the reason of the deviations shown in Fig. \ref{MT} and \ref{AES} is a positive correlation among the test items.

\begin{figure}
    \centering\subfigure[Using  Mersenne twister as the generator.]{\includegraphics[scale=0.25]{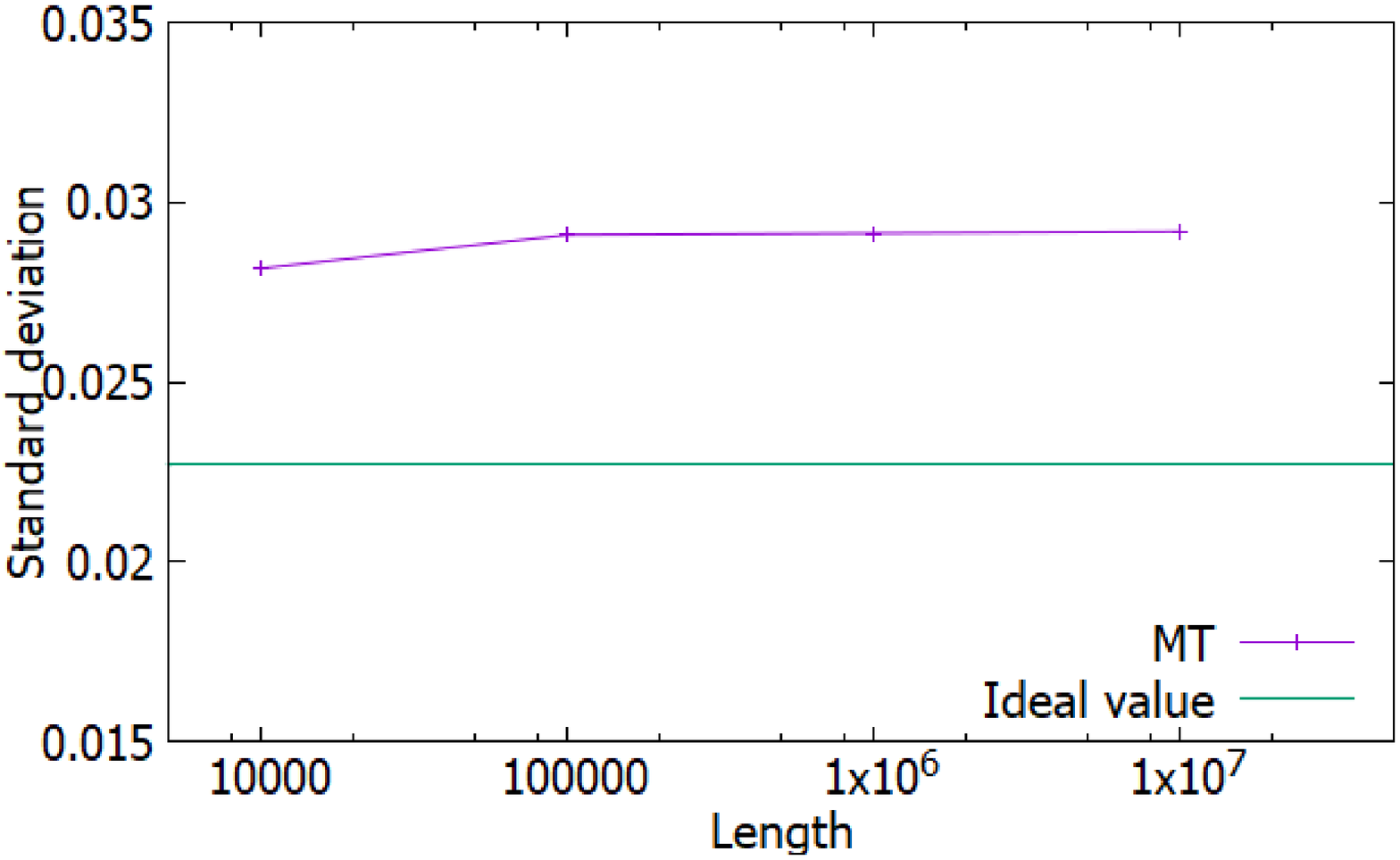}}
    \centering\subfigure[Using  AES as the generator.]{\includegraphics[scale=0.25]{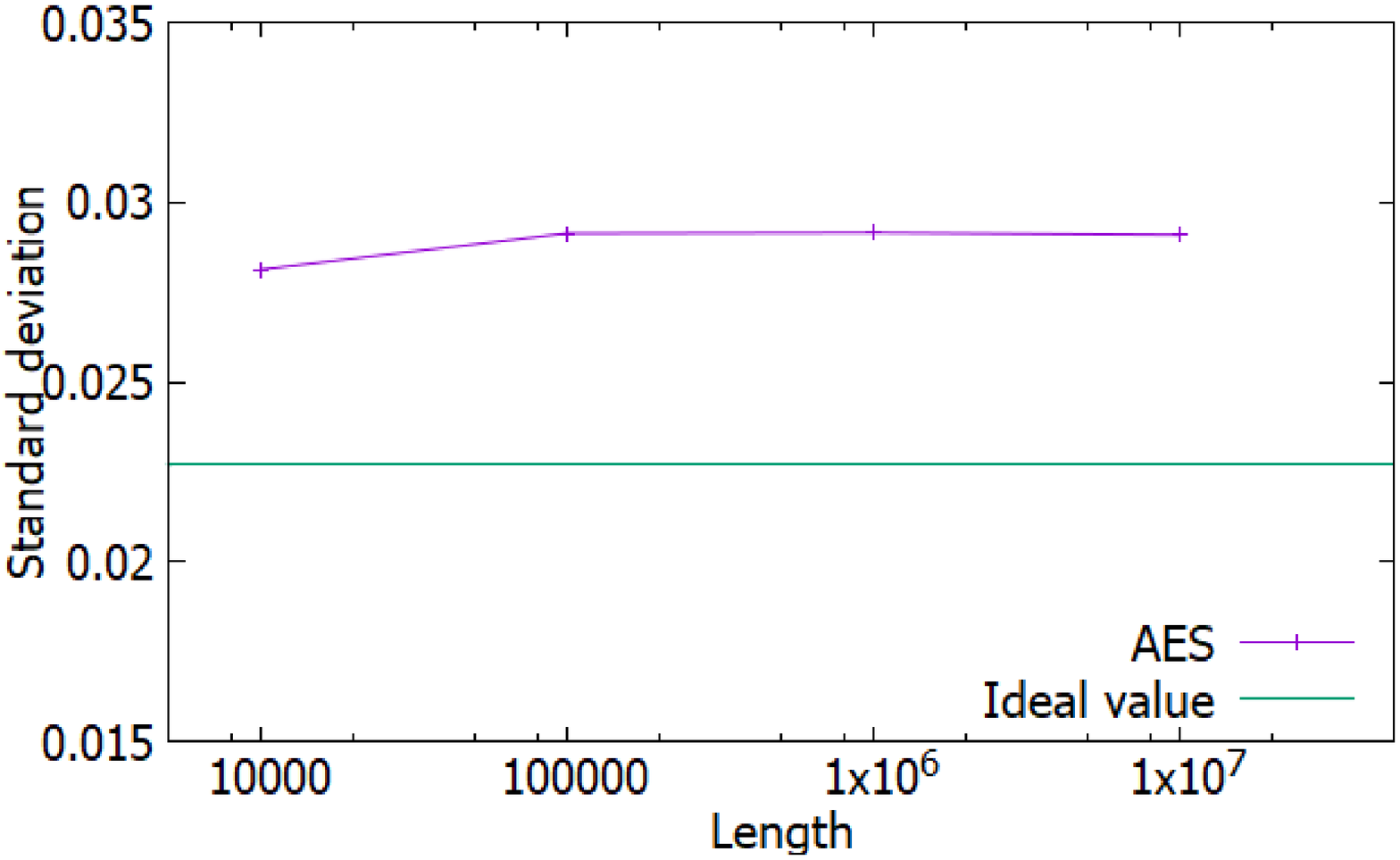}}
\caption{Comparing the standard deviation of the histograms of $q_1,q_2,\cdots,q_m$ with that of the theoretical distribution.}
\label{deviation}
\end{figure}

Fig. \ref{deviation} shows the standard deviations of the histograms made from Fig. \ref{MT} and \ref{AES}.
The author does not know whether the positive correlation among the test items will vanish as $n\to\infty$ or not.
We can at least say that the correlation will not vanish while $n$ is in a range we can realistically deal with.


\section{Deriving a ``minimum'' set}

If there is a strong correlation between two test items, it is enough to perform only one test out of the two.
Then, in order to reduce amount of computation, subsets of test items included in SP800-22 have been proposed as alternatives to the original SP800-22.
These subsets are called ``minimum sets".
Removing test items which have a strong correlation with other test items will be useful for making a rational criterion throughout all test items.
In this sentence, the ``all test items" means all items which are not removed.

In Ref. \cite{IPA}, a minimum set is proposed.
The author, however, thinks that the minimum set has a problem.
It is that a clear criterion to select the subset has not been shown.
In other words, ``subjectivity'' is left.
Since a randomness test is used in evaluation of cryptography, ``subjectivity" should be removed.

Then, we propose an algorithmic method to select a subset.
The approach proposed　in Section 2 shows the dependency of the test items as the deviation between the histogram of  $q_1,q_2,\cdots,q_m$ and the theoretical distribution with the assumption of independency.
Then, we use an indicator which measure the distance between the histogram and the theoretical distribution.
If we choose a subset to make the value of the indicator small, test items included in the subset can be expected to do not have strong correlation to each other. 

We show an example.
The definition of $q$ is the same as that in the former section. 
As the indicator measuring the distance between the histogram and theoretical distribution, we use $I$ defined as
\vspace{1mm}
\[\frac{\text{The standard deviation of the histogram}}{\text{The standard deviation of the theoretical distribution}}.\]
\vspace{1mm}

\noindent In the ideal situation, $I$ must be close to 1.
On the other hand, in the real situation, $I$ is bigger than 1.
Thus, we should make $I$ small as the result of removing test items.
\begin{figure}[b]
    \centering
\includegraphics[scale=0.4]{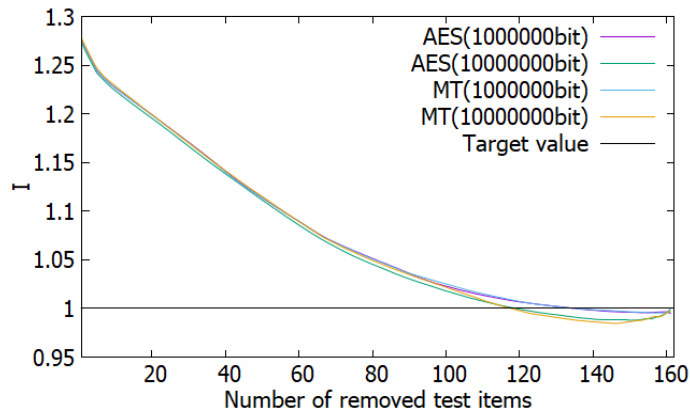}
\caption{Varying of $I$ with removing test items by the greedy algorithm.}
\label{remove}
\end{figure}

It is, however, too difficult to choose the optimal subset because that needs to solve a discrete optimization problem.
Then, we use the greedy algorithm.
Step by step, we choose one test item and remove it to minimize $I$ at that time.

As the result, we got Fig. \ref{remove}.
Although we can see a little bit difference by some conditions, approximately 120-140 items must be removed to make $I$ be close to 1.
Of course, the greedy algorithm is not ensured that it gives the optimal solution and $I=1$ is only a necessary condition that the items which are not removed are mutually independent.
It is, however, a important result because this is first case of such estimation by the scientific method without  subjectivity as far as the author know.

Comparing the last 25 items of the cases  using $10^6$-bit sequences generated by that AES and Mersenne twister, 12 items were included in the both.
If we choose 25 items from 162 items twice, the average of number of items selected twice is approximately 3.9 and the standard deviation is 2.6.
Then, we can say that the result of the proposed method reflects properties of test items to a certain extent.

\section{Conclusion}
The new approach focusing on  randomness of each sequence was proposed.
By the approach, we can visually observe the correlation among the test items included in NIST SP800-22. As the example, we made experiments with the simple average of p-values and actually observed the correlation.

In addition, the method to select a ``minimum set" using the proposed approach was proposed.
By the experiments, we get the conjecture that 120-140 test items included in SP800-22 must be removed to get a subset whose test items are mutually independent.

\end{document}